**Tailoring the magnetic properties of nanocrystalline Cu-Co alloys prepared by high-pressure torsion and isothermal annealing**

A. Bachmaier[1], H. Krenn[2], P. Knoll[2], H. Aboulfadl[3], R. Pippan[1]

[1]Erich Schmid Institute of Materials Science, Austrian Academy of Sciences, Leoben, Austria
[2]Insitute of Physics, University of Graz, Austria
[3]Chair of Functional Materials, Saarland University, Saarbrücken, Germany
Corresponding author: andrea.bachmaier@oeaw.ac.at

**Abstract**

In this study, severe plastic deformation by high-pressure torsion is used as a fabrication method for nanocrystalline magnetic Cu–Co alloys in bulk quantities. By subsequent isothermal annealing, phase separation of the supersaturated solid solutions can be obtained. The magnetic properties of the as-processed and annealed materials have been studied systemically and correlated to the evolving nanostructures investigated in detail by transmission electron microscopy and atom probe tomography. By additional high-pressure torsion deformation at liquid nitrogen temperature the magnetic properties of the $Cu_{74}Co_{26}$ alloy can be further tuned.

# 1. Introduction

The magnetic properties of nanocrystalline Cu-Co alloys have received enormous attention since early studies carried out by Childress and Chien [1] and Berkowitz et al. [2], who synthesized heterogeneous thin film Cu-Co alloys by direct current magnetron sputtering, and Xiao et al. [3], who used vapor quenching to prepare nanogranular Cu-Co materials. Cu and Co are immiscible at room temperature (RT) but metastable solid solutions can be prepared in this alloy system, for example, with the aforementioned methods [1-3]. Subsequent annealing treatments cause decomposition and in the ideal case, nanometer sized ferromagnetic Co particles embedded in the non-magnetic Cu matrix develop. In the past decades, many different techniques, i.e. chemical deposition [4], cluster deposition [5], melt-spinning [6-9], electrodeposition [10-12] and mechanical alloying [13-16], have been examined for production and control of the microstructure of the Cu-Co alloys in the as-processed or subsequent annealed states. A wide variety of microstructures characterized by different size distributions of the precipitated Co particles in the Cu matrix or by different magnetic phases, as a result of a Cu matrix exhibiting a spinodal like Co concentration profile, resulted in various magnetic properties [9,12,17-21]. One major drawback is, however, that all of these techniques are not suitable for mass production. Deposition techniques, for example, produce Cu-Co materials with well controllable structures, but allow one only to obtain rather thin films. Mechanically alloyed Cu-Co powders need to be compacted after processing. Melt-spinning produces long ribbons with reasonable dimensions in two directions, but they are only tens of microns thick.

Severe plastic deformation (SPD) is another way to synthesize massive bulk Cu-Co alloys more conveniently [22]. Subsequent consolidation steps are not needed and materials with reasonable size can be obtained. The effect of SPD on magnetic properties of casted Cu–Co alloys with rather low Co (up to 10 wt.%) or Cu contents (up to 13.6 wt.%) have already been investigated in first studies [23-26]. In the SPD deformed Cu-rich alloys, fragmentation of Co particles were reported, but no complete supersaturation was achieved and only a fraction of Co particles dissolved in the Cu matrix [23,24,26]. In the SPD deformed Co-rich alloy, the supersaturated Co solid solution in the as-cast state decomposed during deformation and a microstructure of almost pure Co grains and Cu precipitates was formed [25].

Using high-pressure torsion (HPT) deformation, full supersaturation with a single-phase structure in several immiscible material systems, including the Cu-Co system, can be achieved as well [27]. These metastable solid solutions are not in the equilibrium state after HPT processing and during subsequent annealing treatments, decomposition occurs. For a $Cu_{74}Co_{26}$ alloy, it has been shown that phase separation proceeds similar to spinodal decomposition on different length scales at intermediate temperatures [28]. Such mechanisms might be also used to tailor optimized nanostructures in the $Cu_{74}Co_{26}$ alloy with respect to the magnetic properties.

In the present study, the magnetic properties of HPT processed supersaturated solid solutions in the $Cu_{74}Co_{26}$ alloy are studied for the first time. By HPT deformation at different temperatures and subsequent isothermal annealing, different nanostructures are adjusted and with that different magnetic properties can be achieved.

## 2. Experimental

Disks of coarse-grained Cu–26at.% Co composite materials, 8 mm in diameter and 0.6 mm in height, were severely deformed by HPT for 25 rotations at RT resulting in an equivalent strain of >200 at a radius of r=1.5 mm. Details of sample production can be found elsewhere [28,29]. Additionally, samples already HPT deformed at RT (for 25 rotations) were repeatedly deformed at liquid nitrogen (LN) temperature with additional 5 rotations. Selected processed samples were also subjected to isothermal heat treatments at 400°C for 1 h, 7 h and 100 h after HPT deformation.

To measure the magnetic properties, hysteresis loops of the as-processed and annealed samples were recorded at 4 K and 300 K with a magnetic field up to 7 Tesla in a SQUID-magnetometer (QUANTUM DESIGN, MPMS-XL-7). Zero-Field-Cooling (ZFC) and field-cooled (FC) curves with an external applied field of 50 Oe were additionally recorded for samples processed at RT and LN temperature up to a temperature of 300 K. All samples for magnetic investigations were cut from the HPT-deformed disks at a distance >1.5 mm from the sample center.

Atom probe tomography (APT) was carried out on as-processed and annealed samples. Specimens were prepared by site-specific sample preparations in a dual-beam focused ion beam/scanning electron microscopy workstation [30]. The atom probe experiments were carried out on a LEAP™3000X-HR CAMECA™ system in laser pulsing mode at a specimen temperature of 60-65 K, a pulse rate of 160 Hz and pulse energies of 0.5 nJ. IVAS 3.6.11 software was used for the reconstruction and visualization of the APT data. Microstructural investigations were performed with a Philips CM12 transmission electron microscope (TEM) operated with 120 kV.

## 3. Results and discussion

The HPT process leads to a significant grain refinement down to grain sizes of ~100 nm and the formation of supersaturated solid solutions up to 26at.% Co in Cu, investigated by TEM and APT [28,29]. The magnetization curve for the as-processed material is shown in Fig. 1a. The hysteresis loop is not saturated even at a magnetic field as high as 7 Tesla and shows small coercivity and remanence. To modify the structure and initiate decomposition, isothermal heat treatments at 400°C corresponding to the peak temperature of the decomposition process [28] were subsequently carried out for 1, 7 and 100 h. The hysteresis loop of the sample annealed at 400° C for 1 h displays a higher magnetization compared to the as-processed sample (Fig. 1a), but the magnetization is still not saturated at high fields

(144 emu/g$_{Co}$ at an applied field of 7 Tesla). By contrast, almost saturated hysteresis loops are obtained for higher annealing times (7 h and 100 h), although the saturation magnetization values of 150 and 152 emu/g$_{Co}$ are still lower than those reported for bulk hcp Co and fcc Co [1,31,32]. In Fig. 1b, a detailed view of the hysteresis at low applied magnetic fields is shown. The coercivity significantly increases from 75 Oe (as-processed) to 125, 165 and 245 Oe for annealing times of 1, 7 and 100 h, respectively. At the same time, the remanence rises from 7 emu/g$_{Co}$ for the as-processed sample to 27, 38 and 42 emu/g$_{Co}$ for the annealed samples.

The Cu$_{74}$Co$_{26}$ alloy has a nanocrystalline face-centered cubic (fcc) structure after HPT deformation at RT, which is confirmed by TEM [28,29]. To link the magnetic properties to microstructural changes on the nanoscale, APT was carried out on as-processed and annealed samples. Fig. 2a shows an APT reconstruction of the as-processed sample together with isoconcentration surfaces (36at.% Co). The local composition, 25.6at.% Co and 74.2at.% Cu [29], correlate quite well with the nominal composition of the alloy. The formed Cu-rich fcc solid solution is, however, not completely homogeneous and Co is still enriched in nm-sized clusters. The size of the Co-rich nanoclusters ranges between few nm to about 18 nm. The Pearson coefficient µ [33], is determined for several sub-volumes with the size of 20 x 20 x 20 nm³ in different reconstructions of the same condition, showing 0.59±0.23 value. A µ close to zero would indicate a random Co solid solution, whereas the much higher µ of the as-processed sample confirms the visible non-random Co distribution in Fig.2a. Annealing at 400°C results in decomposition of the Cu-rich fcc solid solution (also reported in [28]), leading to the formation of two regions, the first region is composed of Cu-Co finely mixed zone (marked with "M") where fine Cu-rich and Co-rich clusters exist, and the second region consists of alternating Cu and Co particles of relatively larger volumes (Fig.2 c,e,g). The 1D-concentration profiles (Fig. 2b,d,f) taken from the cylindrical region of interests (marked "I-III" in Fig. 2) show that the alternating Cu- and Co- particles are pure in composition. With increasing annealing time, the sizes of these particles grow. For analyzing regions marked

"M" selected volumes (20 x 20 x 20 nm³) were chosen and iso-concentration surfaces of Co (30% Co for 1h and 7h annealed samples and 40% Co for the 100h annealed sample, respectively) were created. The corresponding proxigrams in Fig. 2h display the composition across the very fine clusters in these regions for each annealing condition. Proxigrams show the average composition at the proximity of an iso-concentration surface in the normal direction to the chosen surface, providing a concentration profile that accounts for the 3-D nature of the interface [34]. The Co phase in the mixed region contains about 40at.% Co after annealing at 400°C for 1h. With increasing annealing time, the Co content of the Co-rich phase increases. After 7 h and 100 h annealing, it contains about 50at.% and 100at.% Co, respectively. At the same time, the Co content in the Cu-rich phase decreases to about 22at.% Co (7 h) and 15at.% Co (100 h). The Pearson coefficient calculated from selected mixed sub-volumes (20 x 20 x 20 nm³) is 0.89±0.05, 0.96±0.02 and 0.98±0.02 after 1 h, 7 h and 100 h, respectively.

The hysteresis loop for the as-processed sample cannot be saturated in a high applied magnetic field, which is a hint for a breakdown of long range multi-domain magnetic order. As revealed by APT, the as-processed samples are inhomogeneous at the nanoscale. Although Co atoms are diluted into the Cu matrix and hence give no magnetic contribution, Co is still enriched in nanometer-sized clusters in the fcc Cu phase. Nanoclusters with a size smaller or equal to the exchange length of Co (3.4 nm at 300 K [35]) are supposed to form the non-saturated magnetic phase, whereas Co-rich nanoclusters of >3.4 nm size (exemplarily highlighted by isoconcentration surfaces in Fig. 2a) might form a superparamagnetic phase of enhanced magnetic moment inside the Cu-rich matrix. The appearance of small coercivity and remanence in the hysteresis loop of the as-processed sample visible in Fig. 1b might be attributed to larger Co-rich nanoclusters (in the size range >10 nm in the blocked superparamagnetic state). To study the magnetic behavior at low temperature, a hysteresis loop of the as-processed sample was additionally recorded at 4 K (Fig. 3). At this measurement temperature, a nearly saturated hysteresis loop with significant higher magnetization values (i.e., already 133 emu/$g_{Co}$ at 2 Tesla) is obtained (Fig. 3a). The magnetization curve measured at 4 K further shows a distinct hysteresis with a coercivity of ~400 Oe and a remanence of 48 emu/$g_{Co}$ (Fig. 3b), which clearly indicates a dominant blocked superparamagnetic phase at low temperatures.

The critical cluster volume for blocked superparamagnetism (at which magnetocrystalline anisotropy becomes relevant) is directly proportional to the magnetic blocking temperature [36,37] manifested by a peak in the temperature scan of DC susceptibility for a mono-disperse diameter distribution of particles: the lower the blocking temperature, the lower the critical size for blocked superparamagnetic clusters. The Co-rich nanoclusters in the as-processed state have a certain size distribution as displayed by APT. Depending on their size, they either show non-hysteretic superparamagnetic behavior (at temperatures above their blocking temperature for small clusters) or blocked (hysteretic) superparamagnetic behavior (at temperatures below their blocking temperature for larger particles) [37]. Hence, more nanoclusters are supposed to fall into the non-hysteretic superparamagnetic regime if the measurement temperature is higher which is indeed observed in the as-processed sample.

For the annealed samples, the blocked superparamagnetic character of the hysteresis loop becomes dominant at 300 K (Fig. 1a) due to an increase of the cluster size. Since annealing induces phase separation of the Cu-rich solid solution, the amount of Co atoms (which are diluted in the Cu matrix) diminishes and segregate into growing clusters. Near the grain boundary regions, additional pure Co particles are formed which are big enough for the predominance of a blocked superparamagnetic response. At the same time, a collective magnetic phase develops during annealing, arising from the composition modulation in the mixed Cu-Co regions [19-21]. The Co concentration in these mixed regions increases with the annealing time and simultaneously the Co particles grow larger in size. As a consequence, the combined mixed and blocked superparamagnetic phase (in the following loosely denoted as

"ferromagnetic") gain further weight by the evolution of hysteresis. Nevertheless, the measured saturation magnetization value at 7 Tesla is still below the bulk Co value after annealing for 100 h, since the minimum particle size for multi-domain formation (56 nm) is still not exceeded. If all Co atoms would form a percolating network of magnetically coupled spins causing itinerant magnetism the total saturation magnetization would be that of metallic Co [1,31,32]. Hence, some atomic-size Co with diminished magnetic response is still dissolved into the Cu matrix as confirmed by APT.

The hysteresis measurements performed at 300 K further revealed a strong increase in coercivity with increasing annealing time. In single-domain particles magnetization reversal can occur only by coherent rotation, resulting in considerably higher coercivity than that of multidomain particles [38]. The coherence radius of Co for unison-rotation is 17 nm, the maximum size of a Co particle that will be single-domain is 56 nm [35]. Most of the segregated Co particles near the grain boundary regions are far below this critical size and do not form a percolating network, even after 100 h annealing. Thus, they do not support a multidomain behavior in which a smaller reversal magnetic field is needed to obtain zero net magnetization [38] and the steady increase of the coercivity is explained by the growth of the Co particles upon annealing.

The magnetization curve of the as-processed sample exhibits small, but non-zero coercivity and remanence values. Since the hysteretic behavior of the as-processed $Cu_{76}Co_{24}$ alloy is believed to be caused by Co-rich nanoclusters whose size falls below the critical size (10 nm) for blocked superparamagnetism, hysteresis might be further reduced if the size of the nanoclusters can be generally decreased. This means that a greater homogeneity of the deformation-induced supersaturated solid solutions must be obtained. In principle, such a nanostructure might be achieved with more severe HPT deformation at RT, i.e. larger number of rotations. However, very high shear strains would be needed. A more convenient and better defined process might be subsequent HPT deformation at lowered deformation temperatures.

Hence, the HPT deformation of samples already deformed at RT was repeated at LN temperature. This additional deformation leads to further refinement of the microstructure (Fig. 4a). As is the case with the sample HPT deformed at RT, only a fcc structure is visible in the selected area diffraction pattern in TEM. Fig. 4b shows an APT reconstruction of the LN-sample superimposed by Co isoconcentration surfaces to illustrate possible Co-rich clusters, which are drawn at the same values (36at.% Co) as used in Fig. 2a. In contrast to the sample deformed at RT, no large Co-rich nanoclusters are detected. The small Co-rich clusters marked by arrows have a size smaller than 3 nm with much lower distribution density. The Pearson coefficient µ, which is calculated from the whole APT data volume in this case, gives a value of only 0.24 and further confirms the more random Co distribution.

The magnetic properties of the sample deformed at LN temperature are also measured by SQUID. Compared to the sample deformed at RT, magnetization is smaller by a factor of 1.2 and also does not saturate at 300K (Fig. 5a). Two factors might contribute to the reduced magnetization: solid solution formation as observed in the sample deformed at RT as well, and a further reduction in the number of Co-Co nearest neighbors [39]. Radial distribution function analysis (RDF) of the APT data has been applied to study Co clustering effects in the as-processed states (Fig. 4c). In the RDF analysis, the Co concentration surrounding each Co atom is calculated in spherical shells in the 3D space [34], which was then divided by the average bulk concentration. The LN processed sample shows a RDF curve which is close to zero. The RT processed sample, however, has a very small positive interaction at short distances that already indicate a small Co-Co clustering tendency. This collective interaction further dominates in the annealed samples and increases with annealing time. Note also, that RDF is in any case restricted to below the exchange length of Co (3.4 nm), thus collective magnetism with pronounced hysteretic properties is quenched.

The RDF analysis indicates that only extremely fine Co clustering occurs in the sample deformed at LN temperature. In addition, the coercivity and remanence radically decrease to only 7.5 Oe and 1 emu/g$_{Co}$, respectively (Fig. 5b). The very low coercivity and remanence might be either due to the presence of a very small fraction of blocked particles with a size still exceeding the critical size for blocked superparamagnetism or due to the evolution of stress induced anisotropy in small Co-rich nanoclusters [17]. In contrast, magnetic blocking of clusters become the dominant contribution at low measurement temperatures of 4 K. At the same time, the coercivity and remanence increases to 80 Oe and 17 emu/g$_{Co}$ (Fig. 5b). Hysteresis will appear when the particles of a certain size are cooled below the blocking temperature [36,37]. To obtain more precise information about the blocking temperature of the samples processed at RT and LN temperature, ZFC and FC curves were additionally measured (Fig. 5c). The FC and ZFC curve of the sample processed at RT are split up to the highest measurement temperature (300 K) indicating a blocking temperature higher than 300 K. By contrast, the curves of the sample processed at LN temperature converge below 300 K and a blocking temperature of about 275 K is observed. The better homogeneity of the deformation-induced solid solution and concomitant smaller Co-rich cluster size leads to an increased fraction of Co-rich nanoclusters which are in the deblocked superparamagnetic regime at 300 K in this sample. At 4 K, however, the major part of the Co-rich nanoclusters belongs to the blocked superparamagnetic regime.

## 4. Conclusion

The study shows that HPT deformation is an alternative fabrication method for magnetic Cu-Co alloys. In the as-processed state, evidence for both, superparamagnetic and ferromagnetic-like behavior, is present. Combining magnetic measurements with results from APT further reveals that it is possible to control hysteretic or non-hysteretic evolution of magnetism from Co-rich nanoclusters in the fcc Cu solid solution. The coercivity, for example, decreases by 10 times due to additional HPT deformation at LN temperature. The magnetic properties of the alloy can thus be also used as a valuable fingerprint for the microstructural changes. A drawback is the reduction of saturation magnetization due to the presence of isolated nanoclusters and the absence of a percolating network of connected Co-clusters (in chains etc.) or even domains. The magnetization dynamics is different from the conventional domain-wall motion. However, precipitations of magnetic clusters in a nanostructured (nonmagnetic) environment obtained by SPD offer new possibilities to tailor magnetic materials in bulky quantities.

**Acknowledgments**

A. Bachmaier gratefully acknowledges the financial support by the Austrian Science Fund (FWF): J3468-N20. The atom probe instrument was financed by the DFG and the Federal State Government of Saarland (INST 256/298-1 FUGG).

**Figures**

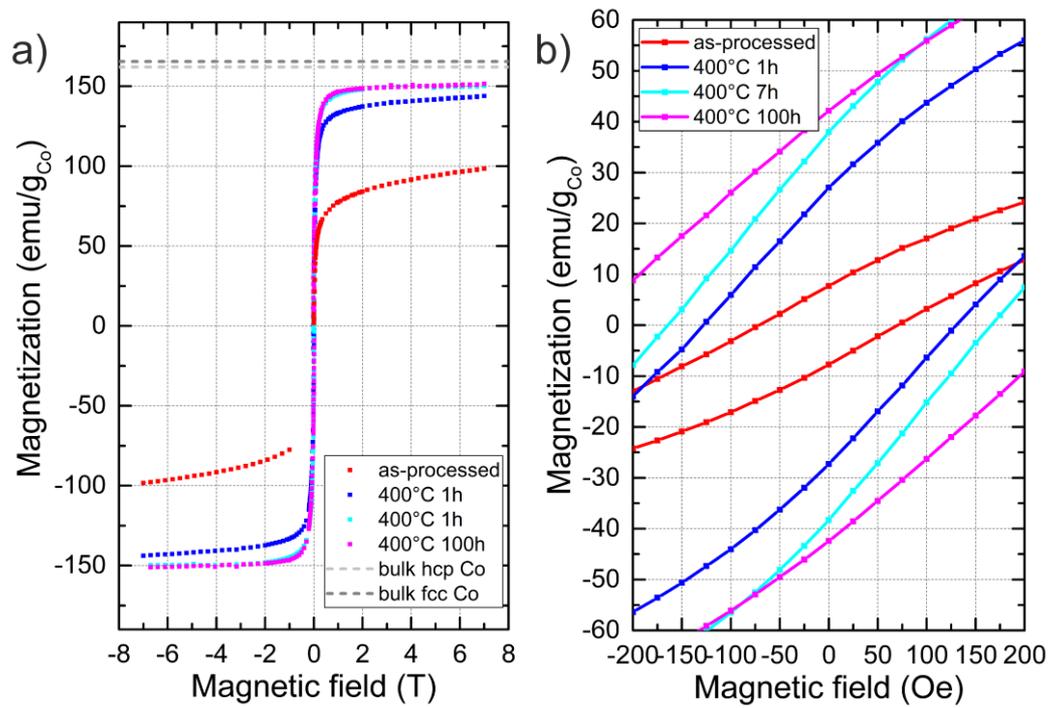

Fig. 1: (a) Hysteresis loops of as-processed (RT) and annealed samples measured at 300 K. (b) Magnified view of the narrow hysteresis.

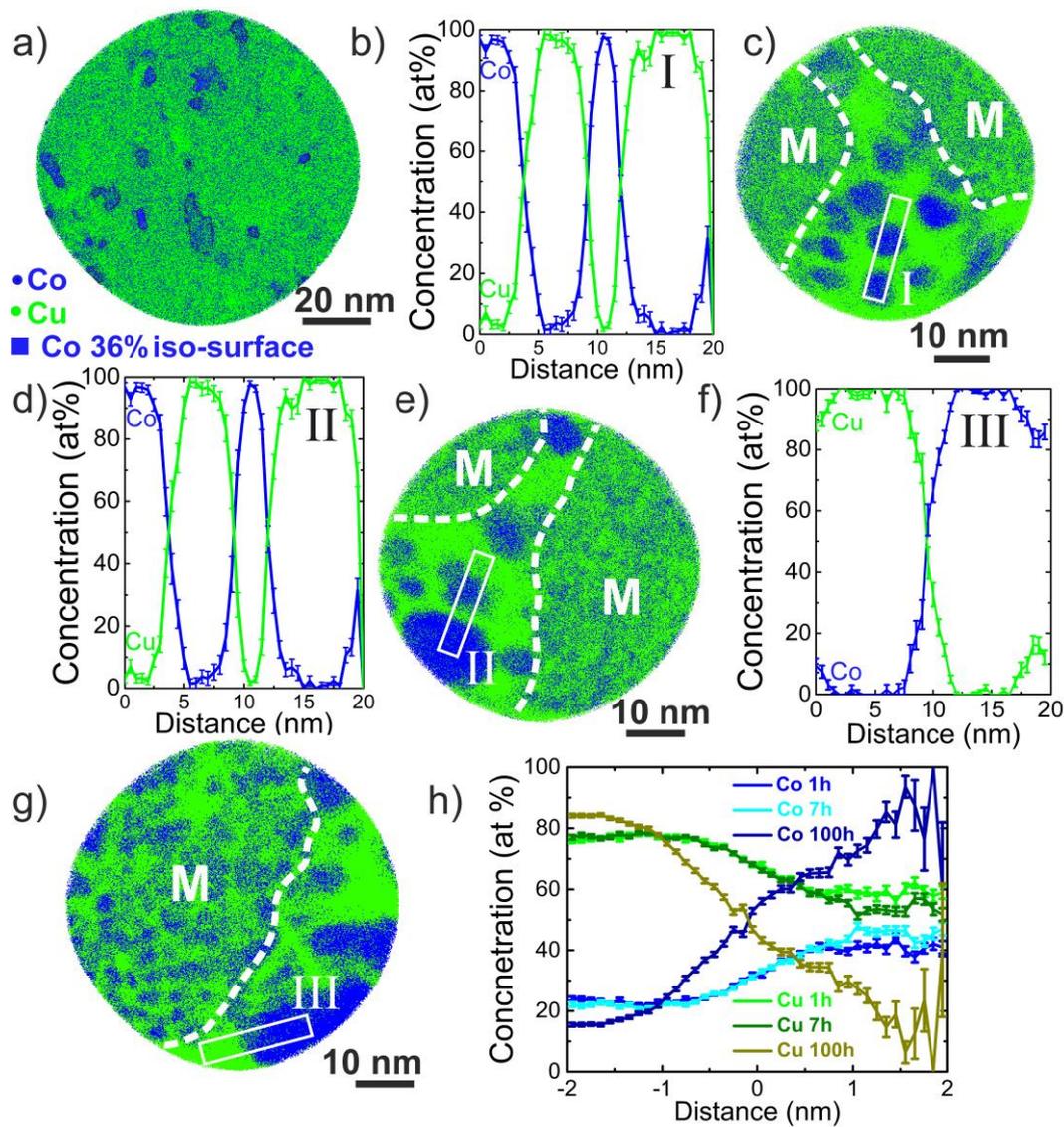

Fig. 2: APT reconstructions for the as-processed sample deformed at RT (a) and for samples annealed at 400°C for 1 h (c), 7 h (e) and 100 h (g). The tomographic slices (top-view) have a thickness of 10 nm, the blue iso-concentration surfaces in (a) correspond to 36at.% Co. (b)-(f) 1-D concentration profiles along the marked region-of-interests (size of sampling cylinders: Ø3 nm x 20 nm). In (h), proxigrams calculated from Co iso-concentration surfaces of selected sub-volumes (20x20x20nm²) are shown (For interpretation of the references to colour in this figure legend, the reader is referred to the web version of this article).

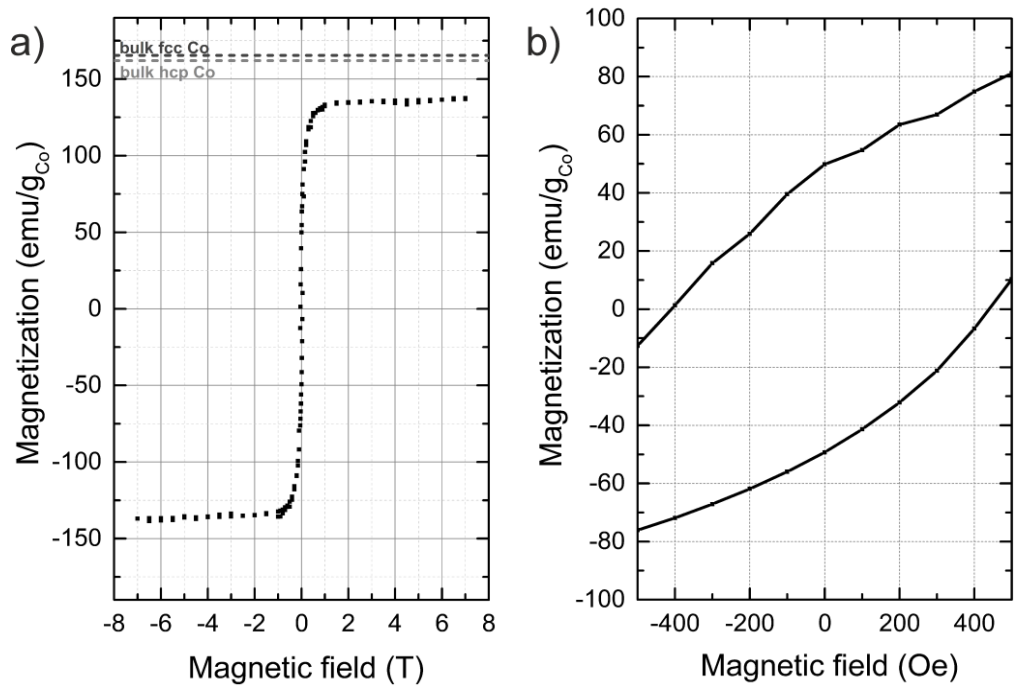

Fig. 3: (a) Hysteresis loops of as-processed (RT) sample measured at 4 K and (b) magnified view of the narrow hysteresis.

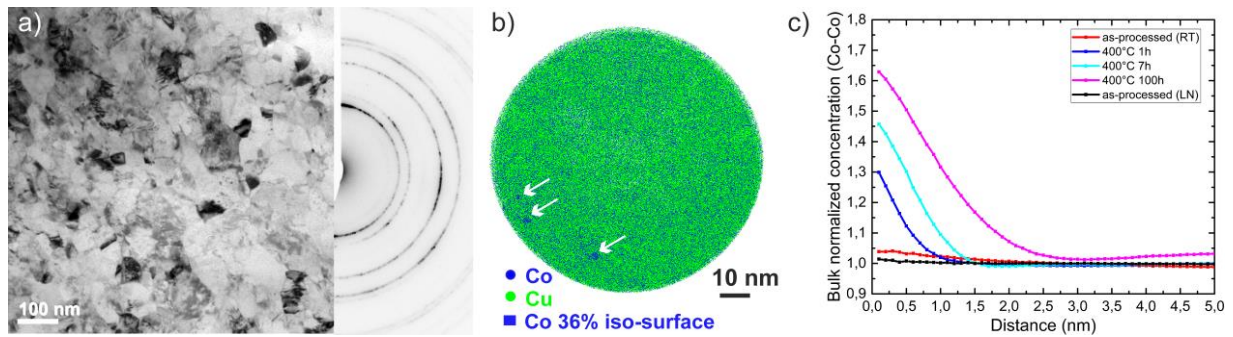

Fig. 4: (a) TEM micrograph and selected area diffraction pattern of as-processed samples with additional deformation at LN temperature. (b) APT reconstruction (tomographic slice in top-view with a thickness of 10 nm) of the same sample. The blue iso-concentration surfaces (marked by white arrows) correspond to 36at.% Co. (c) RDFs of Co-Co for all states (0.1 nm step size). For the LN sample, the RDF is calculated for the whole dataset, for all other samples sub-volumes (20x20x20nm²) inside the mixed regions are chosen (For interpretation of the references to colour in this figure legend, the reader is referred to the web version of this article).

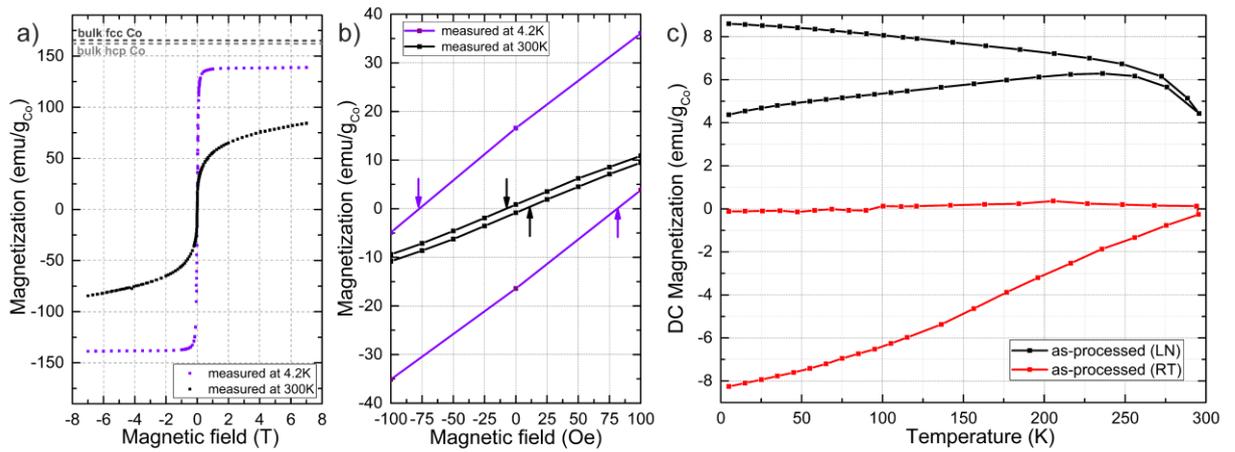

Fig. 5: (a) Hysteresis loops of the as-processed samples deformed at LN temperature measured at 4.2K and 300K. (b) Detailed view of the hysteresis. (c) ZFC and FC curves for the as-processed samples deformed at RT and LN temperature measured with an applied field of 50 Oe up to a temperature of 300 K.